\documentclass[11pt]{article}
\usepackage[colorlinks=true,citecolor=blue,linkcolor=blue]{hyperref}
\usepackage{multirow}% for tables
\usepackage[left=2cm,right=2cm,top=2cm,bottom=2cm]{geometry}
\usepackage{cite}
\usepackage{psfrag}
\usepackage[demo]{graphicx}
\usepackage{graphicx}
\usepackage{graphics}
\usepackage{epsfig}
\usepackage{amsmath,amsfonts,amssymb}
\usepackage{pstcol,pst-fill,pst-grad}
\usepackage{pstricks,pst-fill,pst-grad}
\usepackage{euscript}
\usepackage{pstricks}
\usepackage{wrapfig}
\usepackage{slashed}
\usepackage{here}
\begin{document}
%\newcommand\redsout{\bgroup\markoverwith{\textcolor{red}{\rule[0.5ex]{4pt}{1pt}}}\ULon}
	%\begin{titlepage}
	\title{\vspace{-3cm}
		\hfill\parbox{4cm}{\normalsize \textit{}}\\
		\vspace{1cm}
		{Analysis of the geometric effect on laser-assisted decay processes}}
	\vspace{2cm}
	
	\author{S Mouslih$^{2,1}$, M Jakha$^{1}$, S El Asri$^{1}$, S Taj$^1$, B Manaut$^{1,}$\thanks{Corresponding author, E-mail: b.manaut@usms.ma} and E Siher$^{2}$ \\
		{\it {\small$^1$ Sultan Moulay Slimane University, Polydisciplinary Faculty,}}\\
		{\it {\small Laboratory of Research in Physics $\&$ Engineering Sciences, Team of Modern and Applied Physics,}}\\
		{\it {\small Beni Mellal, 23000, Morocco.}}
		\\			
	{\it {\small$^2$Faculty of Sciences and Techniques, 
		Laboratory of Materials Physics (LMP),
		Beni Mellal, 23000, Morocco.}}		
	}
	\maketitle \setcounter{page}{1}



%\date{\today}

\begin{abstract}
Choosing a specific direction for the propagation of laser field waves often presents a challenge for researchers studying laser-assisted ultrafast quantum processes. They are faced with the question of why exactly this direction and not another. This paper resolves the discussion in this issue regarding decay processes. Therefore, we study theoretically the pion decay process in the presence of a circularly polarized laser field propagating along an arbitrary general direction. Using the first Born approximation and the Dirac-Volkov states for charged particles, we derive an analytic expression for the decay rate. The direction of the laser field was found to have no significant effect on the nature of the result obtained. This study generalizes the results found for a field with a wave vector along the $z$-axis in a recent paper (Phys Rev D 102:073006, 2020). This paper will serve as a justification for the choice of a specific direction for the laser field in laser-assisted decay processes. The effect of the laser field on the total decay rate has also been reported and discussed.
\end{abstract}
Keywords: laser-assisted; electroweak processes; Dirac-Volkov state; decay rate 

\maketitle
%\newpage
\section{Introduction}
The study of quantum processes in the presence of laser field is a fertile area of research that has attracted the attention of many theoretical and experimental scientists \cite{piazza}, due to the development of laser technology \cite{laser}. These studies contribute significantly to understanding the laser-matter interaction. The ultimate goal behind all this is to understand the behavior of particles and to discover their new properties arising in the presence of an external field. The ultrafast processes that occur in the presence of the laser field are varied, depending on the adopted framework of study. In atomic physics, many atomic processes have been studied in the presence of an electromagnetic field, both in relativistic and non-relativistic regimes \cite{manaut1,manaut2,manaut3,manaut4,manaut5,Roshchupkin1,Roshchupkin2}. In the framework of quantum electrodynamics and electroweak theory, many articles have been devoted to the investigation of scattering \cite{Roshchupkin3,dahiri,ouhammou1,ouhammou2,mouha} and decay \cite{mouslih,jakha,baouahi,wdecay} processes in the presence of an electromagnetic field. All these studies use a specific direction along which the laser field propagates. In this context, we have recently studied the pion decay process in the presence of a circularly polarized electromagnetic field with a wave vector along the $z$-axis direction \cite{mouslih}. We found that the laser contributed to the increase of the lifetime, and this result was explained by the so-called quantum Zeno effect. In this paper, we will extend this study to the case of general laser field direction. This would lead to a generalization of the results obtained previously. This is done by configuring the wave vector with a general spherical geometry that allows us to cover all particular cases. The main purpose of this work is to establish a general theoretical formalism that allows us to examine the effect of laser field direction on different measurable quantities in decay processes. More details on the theoretical calculation and the literature on this subject can be found in our previous article \cite{mouslih}. Throughout this work, we use natural units $c=\hbar=1$ and the metric tensor $g=diag(1,-1,-1,-1)$.
\section{Theoretical formalism}
The process considered is the decay of a charged pion $\pi^{-}$ into two leptons,
\begin{equation}\label{process}
 \pi^{-}(p_{1})\longrightarrow \ell^{-}(p_{2})+\bar{\nu}_{\ell}(k'), ~~~~~~(\ell=e,\mu)
\end{equation}
where the arguments are our labels for the associated four-momenta. The laser field is assumed to be monochromatic and circularly polarized. Its classical four-potential satisfies the Lorentz gauge condition $\partial_{\mu}A^{\mu}=0,$ and is given by:
\begin{equation}\label{potential}
A^{\mu}(\phi)=a^{\mu}_{1}\cos(\phi)+a^{\mu}_{2}\sin(\phi),
\end{equation}
where $\phi=(k.x)$ is the phase of the laser field. The wave 4-vector $k$ is introduced theoretically in a general geometry with spherical coordinates as follows:
\begin{equation}\label{wavevector}
k=(\omega,\textbf{k})=\omega\Big(1,\cos(\varphi_k)\sin(\theta_k),\sin(\varphi_k)\sin(\theta_k),\cos(\theta_k)\Big),
\end{equation}
where $\omega$ is the frequency of the laser field. The Lorentz gauge condition applied to the four-potential $A^{\mu}$ implies that $k_{\mu}A^{\mu}=0$, meaning that $(k.a_{1})=(k.a_{2})=0$. To keep these relationships verified, we set also the polarization 4-vectors $a^{\mu}_{1}$ and $a^{\mu}_{2}$ in a general spherical geometry as follows:
\begin{equation}
\begin{split}
a^{\mu}_{1}&=(0,\mathbf{a}_{1})=|\mathbf{a}|\Big(0,\cos(\varphi_{a_{1}})\sin(\theta_{a_{1}}),\sin(\varphi_{a_{1}})\sin(\theta_{a_{1}}),\cos(\theta_{a_{1}})\Big),\\
a^{\mu}_{2}&=(0,\mathbf{a}_{2})=|\mathbf{a}|\Big(0,\cos(\varphi_{a_{2}})\sin(\theta_{a_{2}}),\sin(\varphi_{a_{2}})\sin(\theta_{a_{2}}),\cos(\theta_{a_{2}})\Big),
\end{split}
\end{equation}
with $|\mathbf{a}|=\mathcal{E}_{0}/\omega$, where $\mathcal{E}_{0}$ is the laser field strength. The polarization 4-vectors are orthogonal and equal in magnitude, which implies $(a_{1}.a_{2})=0$ and $a_{1}^{2}=a_{2}^{2}=a^{2}=-|\mathbf{a}|^2$.\\
The wave function of the relativistic lepton $\ell^{-}$, with four-momentum $p_{2}$ and spin $s_{1}$, moving in an electromagnetic field was first presented by Volkov in 1935 \cite{volkov}, and is given, when normalized to the volume $V$, by \cite{landau}
\begin{equation}\label{muon}
\psi_{\ell}(x)=\bigg[1+\frac{e\slashed{k}\slashed{A}}{2(k.p_{2})}\bigg]\frac{u(p_{2},s_{1})}{\sqrt{2Q_{2}V}}\times e^{iS(q_{2},x)},
\end{equation}
where $e=-|e|$ is the charge of the electron, and 
\begin{equation}
S(q_{2},x)=-q_{2}.x-\frac{e(a_{1}.p_{2})}{k.p_{2}}\sin(\phi)+\frac{e(a_{2}.p_{2})}{k.p_{2}}\cos(\phi),
\end{equation}
with the electron's effective momentum and mass
\begin{equation}
q_{2}=p_{2}-\frac{e^{2}a^{2}}{2(k.p_{2})}k, ~~~m_{\ell*}^{2}=m_{\ell}^{2}-e^{2}a^{2},
\end{equation}
where $m_{\ell}$ is the rest mass of the free lepton.\\
For the laser-dressed charged pion (spinless particle), its wave function obeys the Klein-Gordon equation for bosons with spin zero. Therefore, the corresponding Volkov solution reads \cite{szymanowski}:
\begin{equation}\label{pion}
\psi_{\pi^{-}}(x)=\frac{1}{\sqrt{2Q_{1}V} }
\times e^{iS(q_{1},x)},
\end{equation} 
with
\begin{equation}
S(q_{1},x)=-q_{1}.x-\frac{e(a_{1}.p_{1})}{k.p_{1}}\sin(\phi)+\frac{e(a_{2}.p_{1})}{k.p_{1}}\cos(\phi).
\end{equation}
The outgoing antineutrino $\bar{\nu}_{\ell}$ is treated as massless particle with four-momentum $k'$ and spin $s_{2}$. Its wave function is given by \cite{greiner}:
\begin{equation}
\psi_{\bar{\nu}_{\ell}}(x)=\frac{v(k',s_{2})}{\sqrt{2E_{2}V}  }e^{ik'.x},
\end{equation}
where $E_{2}=k'^{0}$ is the total energy of the outgoing antineutrino.\\
In the first Born approximation, the S-matrix  element for the laser-assisted pion decay can be written as \cite{greiner}:
\begin{equation}\label{smatrix}
S_{fi}(\pi^{-}\rightarrow \ell^{-}\bar{\nu}_{\ell})=\frac{-iG}{\sqrt{2}}\int d^{4}xJ^{(\pi)\dagger}_{\mu}(x)J^{\mu}_{(\ell^{-})}(x),
\end{equation}
where $G$ is the Fermi coupling constant. $J^{\mu}_{(\ell^{-})}(x)$ and $J^{(\pi)}_{\mu}(x)$ are, respectively, the leptonic and hadronic currents expressed by:
\begin{equation}\label{currents}
\begin{split}
J^{\mu}_{(\ell^{-})}(x)&=\bar{\psi}_{\ell}(x,t)\gamma^{\mu}(1-\gamma_{5})\psi_{\bar{\nu}_{\ell}}(x,t),\\
J^{(\pi)}_{\mu}&=i\sqrt{2}f_{\pi}p_{1\mu}\frac{1}{\sqrt{2Q_{1}V} }
\times e^{-iS(q_{1},x)},
\end{split}
\end{equation}
where $f_{\pi}=90.8~\mbox{MeV}$ is called the pion decay constant \cite{greiner}.\\
Inserting the expressions of currents and wave functions into Eq.~(\ref{smatrix}) and after some algebraic manipulations, we get:
\begin{equation}
S_{fi}=\frac{-Gf_{\pi}}{2\sqrt{2Q_{1}Q_{2}E_{2}V^{3}}}\sum_{n=-\infty}^{\infty}M^{n}_{fi}(2\pi)^{4}\delta^{4}(k'+q_{2}-q_{1}-nk),
\end{equation}
where $n$ is the number of exchanged photons, and the quantity $M^{n}_{fi}$ is defined by:
\begin{equation}
M^{n}_{fi}=\bar{u}(p_{2},s_{1})\Gamma^{n}v(k',s_{2}),
\end{equation}
where
\begin{equation}
\Gamma^{n}=\Big\lbrace B_{n}(z)+ [e/(2 (k.p_{2}))] \slashed{a}_{1}\slashed{k} B_{1n}(z)+[e/(2 (k.p_{2}))] \slashed{a}_{2}\slashed{k} B_{2n}(z)\Big\rbrace\slashed{p}_{1}\Big(1-\gamma_{5}\Big).
\end{equation}
The coefficients $B_{n}(z)$, $B_{1n}(z)$ and $B_{2n}(z)$ are expressed in terms of ordinary Bessel functions by:
\begin{equation}
\left[\begin{array}{c}
B_{n}(z)\\
B_{1n}(z)\\
B_{2n}(z) \end{array}\right]=\left[\begin{array}{c} J_{n}(z)e^{in\phi_{0}}\\
\big(J_{n+1}(z)e^{i(n+1)\phi_{0}}+J_{n-1}(z)e^{i(n-1)\phi_{0}}\big)/2\\
\big(J_{n+1}(z)e^{i(n+1)\phi_{0}}-J_{n-1}(z)e^{i(n-1)\phi_{0}}\big)/2i
 \end{array}\right],
\end{equation}
where $\phi_{0}=\arctan(\alpha_{2}/\alpha_{1})$, and the argument of the Bessel function $z$ is defined by:
\begin{equation}\label{argument}
 z=\sqrt{\alpha_{1}^{2}+\alpha_{2}^{2}}\quad\mbox{with}\quad\alpha_{1}=e\bigg(\frac{a_{1}.p_{1}}{k.p_{1}}-\frac{a_{1}.p_{2}}{k.p_{2}}\bigg)\,;\,\alpha_{2}=e\bigg(\frac{a_{2}.p_{1}}{k.p{1}}-\frac{a_{2}.p_{2}}{k.p_{2}}\bigg).
\end{equation}
Following the same procedure as detailed in \cite{mouslih}, we obtain for the decay rate:
\begin{equation}\label{summed}
W(\pi^{-}\rightarrow \ell^{-}\bar{\nu}_{\ell})=\sum_{n=-\infty}^{+\infty}W_{n}(\pi^{-}\rightarrow \ell^{-}\bar{\nu}_{\ell}),
\end{equation}
where the $n$-resolved decay rate $W_{n}$ is defined by: 
\begin{equation}\label{ws}
W_{n}=\frac{G^{2}f_{\pi}^{2}}{(2\pi)^{2}8Q_{1}}\int \frac{|\mathbf{q}_{2}|^{2}d\Omega_{\ell}}{E_{2}Q_{2}g'(|\mathbf{q}_{2}|)}|\overline{M^{n}_{fi}}|^{2},
\end{equation}
with
\begin{equation}
g'(|\mathbf{q}_{2}|)=\frac{|\mathbf{q}_{2}|-n\omega\cos(\theta)}{\sqrt{(n\omega)^{2}+|\mathbf{q}_{2}|^{2}-2n\omega|\mathbf{q}_{2}|\cos(\theta)}}+\frac{|\mathbf{q}_{2}|}{\sqrt{|\mathbf{q}_{2}|^{2}+m_{\ell*}^{2}}},
\end{equation}
where $\theta$ is the final angle of the emitted lepton. The term  $|\overline{M^{n}_{fi}}|^{2}=\sum_{s_{1},s_{2}}|M^{n}_{fi}|^{2}$ in Eq.~(\ref{ws}) reduces down to the following trace:
\begin{equation}\label{trace}
|\overline{M^{n}_{fi}}|^{2}=\mbox{Tr}\big[(\slashed{p}_{2}+m_{\ell})\Gamma^{n}\slashed{k}'\bar{\Gamma}^{n}\big],
\end{equation}
where
\begin{equation}
 \begin{split}
\bar{\Gamma}^{n}&=\gamma^{0}\Gamma^{n\dagger}\gamma^{0},\cr
&=\slashed{p}_{1}\Big(1-\gamma_{5}\Big)\Big\lbrace B^{*}_{n}(z)+ [e/(2 (k.p_{2}))]\slashed{k}\slashed{a}_{1}B^{*}_{1n}(z)+[e/(2 (k.p_{2}))] \slashed{k}\slashed{a}_{2}B^{*}_{2n}(z)\Big\rbrace.
\end{split}
\end{equation}
The trace work can be done numerically with the help of FEYNCALC \cite{feyncalc}. The result we have obtained is as follows:
\begin{equation}\label{spinorpart}
\begin{split}
|\overline{M^{n}_{fi}}|^{2}=\frac{4}{(k.p_{2})^2}\Big[&\Delta_{1}|B_{n}|^{2}+\Delta_{2}|B_{1n}|^{2}+\Delta_{3}|B_{2n}|^{2}+\Delta_{4}B_{n}B^{*}_{1n}+\Delta_{5}B_{1n}B^{*}_{n}+\Delta_{6}B_{n}B^{*}_{2n}\cr
&+\Delta_{7}B_{2n}B^{*}_{n}+\Delta_{8}B_{1n}B^{*}_{2n}+\Delta_{9}B_{2n}B^{*}_{1n}\Big],
\end{split}
\end{equation}
where the nine coefficients from $\Delta_{1}$ to $\Delta_{9}$ are explicitly expressed, in terms of different scalar products, by
\begin{equation}
\Delta_{1}=4 (k.p_{2})^2 (p_{1}.p_{2}) (p_{1}.k')- 2(k.p_{2})^2 m_{\pi^{-}}^2 (p_{2}.k'),
\end{equation}
\begin{equation}
\Delta_{2}=-e^2 \big(2 (a_1.k) (a_{1}.p_{2})- a^2 (k.p_{2})\big) \big((k.k') m_{\pi^{-}}^2 - 2 (k.p_{1})(p_{1}.k')\big),
\end{equation}
\begin{equation}
\Delta_{3}=-e^2\big(2 (a_{2}.k) (a_{2}.p_{2}) - a^2 (k.p_{2}) \big)\big((k.k') m_{\pi^{-}}^2 - 2 (k.p_{1})(p_{1}.k') \big),
\end{equation}
\begin{equation}
\begin{split}
\Delta_{4}=&e (k.p_{2}) \big(2 (a_{1}.p_{2}) (k.p_{1}) (p_{1}.k') + 2 (a_{1}.k) (p_{1}.p_{2}) (p_{1}.k') \\ &- (a_{1}.k) m_{\pi^{-}}^2 (p_{2}.k')\big),
\end{split}
\end{equation}
\begin{equation}
\begin{split}
\Delta_{5}=&e (k.p_{2})\big[(a_{1}.k')(k.p_{2})  m_{\pi^{-}}^2 - (a_{1}.p_{2})(k.k')  m_{\pi^{-}}^2 + 2 (a_{1}.p_{2})\\
&\times(k.p_{1})(p_{1}.k') + 2 (a_{1}.k)(p_{1}.p_{2}) (p_{1}.k') - (a_{1}.k)  m_{\pi^{-}}^2 (p_{2}.k')\cr&-i\big(2 (p_{1}.k') \epsilon(a_{1},k,p_{1},p_{2}) 
 + m_{\pi^{-}}^2 \epsilon(a_{1},k,p_{2},k')
\big)\big],
\end{split}
\end{equation}
\begin{equation}
\begin{split}
\Delta_{6}=&e (k.p_{2})\big[(a_{2}.k') (k.p_{2}) m_{\pi^{-}}^2 - (a_{2}.p_{2}) (k.k') m_{\pi^{-}}^2 + 2 (a_{2}.p_{2}) (k.p_{1})\cr&\times (p_{1}.k') +2 (a_{2}.k)(p_{1}.p_{2})
(p_{1}.k')- (a_{2}.k) m_{\pi^{-}}^2 (p_{2}.k')-i\big(2 (p_{1}.k')\cr&\times\epsilon(a_{2},k,p_{1},p_{2}) 
 + m_{\pi^{-}}^2 \epsilon(a_{2},k,p_{2},k')
\big)\big],
\end{split}
\end{equation}
\begin{equation}
\begin{split}
\Delta_{7}=&e (k.p_{2})\big[(a_{2}.k')(k.p_{2}) m_{\pi^{-}}^2 - (a_{2}.p_{2})(k.k') m_{\pi^{-}}^2 + 2 (a_{2}.p_{2})\cr
&\times(k.p_{1})(p_{1}.k')+2 (a_{2}.k)(p_{1}.p_{2}) (p_{1}.k')- m_{\pi^{-}}^2(a_{2}.k) (p_{2}.k')\cr&+2 i (p_{1}.k') \epsilon(a_{2},k,p_{1},p_{2}) + i m_{\pi^{-}}^2 \epsilon(a_{2},k,p_{2},k')\big],
\end{split}
\end{equation}
\begin{equation}
\begin{split}
\Delta_{8}=&e^2\big[-\big((a_{1}.p_{2})(a_{2}.k) + (a_{1}.k)(a_{2}.p_{2})-(a_{1}.a_{2}) (k.p_{2})\big)\big((k.k') m_{\pi^{-}}^2 \cr&- 2 (k.p_{1})(p_{1}.k')\big)+i \big(2 (a_{1}.k)(p_{1}.k')\epsilon(a_{2},k,p_{1},p_{2}) 
- (a_{2}.k) m_{\pi^{-}}^2 \cr &\times \epsilon(a_{1},k,p_{2},k')+ (a_{1}.k) m_{\pi^{-}}^2 \epsilon(a_{2},k,p_{2},k') \big)\big],
\end{split}
\end{equation}
\begin{equation}
\begin{split}
\Delta_{9}=&e^2\big[-\big((a_{1}.p_{2})(a_{2}.k) + (a_{1}.k)(a_{2}.p_{2})-(a_{1}.a_{2}) (k.p_{2})\big)\big((k.k') m_{\pi^{-}}^2 \cr&- 2 (k.p_{1})(p_{1}.k')\big) -i\big(2(k.p_{2})(p_{1}.k') \epsilon(a_{1},a_{2},k,p_{1})-2(a_{2}.k) \cr&\times (p_{1}.k')\epsilon(a_{1},k,p_{1},p_{2})+ 2 (a_{1}.k)(p_{1}.k')\epsilon(a_{2},k,p_{1},p_{2})\cr&- (k.p_{2}) m_{\pi^{-}}^2 \epsilon(a_{1},a_{2},k,k')
- (a_{2}.k)  m_{\pi^{-}}^2 \epsilon(a_{1},k,p_{2},k')\cr&+ (a_{1}.k) m_{\pi^{-}}^2 \epsilon(a_{2},k,p_{2},k') 
\big)\big],
\end{split}
\end{equation}
where the different scalar products are evaluated in the rest frame of pion; and for all 4-vectors $a, b, c$ and $d$, we have
\begin{equation}
\epsilon(a,b,c,d)=\epsilon^{\mu\nu\rho\sigma}a_{\mu}b_{\nu}c_{\rho}d_{\sigma},
\end{equation}
where $\epsilon^{\mu\nu\rho\sigma}$ is the antisymmetric tensor with the convention $\epsilon^{0123}=1$. The reader may refer to our previous works \cite{mouslih,jakha} to see how these tensors are calculated analytically.\\
After the decay rate, it comes the lifetime of the charged pion, which is defined simply as the inverse of the total decay rate
\begin{equation}\label{pionlifetime}
\tau_{\pi^{-}}=\frac{1}{W_{total}},~~~\mbox{with}~~W_{total}=W(\pi^{-}\rightarrow \mu^{-}\bar{\nu}_{\mu})+W(\pi^{-}\rightarrow e^{-}\bar{\nu}_{e}).
\end{equation}
Considering the two decay modes (muonic and electronic) of the charged pion, we define their corresponding branching ratios (Br) as follows:
\begin{eqnarray}
\mbox{Br}(\pi^{-}\rightarrow \mu^{-}\bar{\nu}_{\mu})&=&\frac{W(\pi^{-}\rightarrow \mu^{-}\bar{\nu}_{\mu})}{W_{total}},\label{brmuon}\\
\mbox{Br}(\pi^{-}\rightarrow e^{-}\bar{\nu}_{e})&=&\frac{W(\pi^{-}\rightarrow e^{-}\bar{\nu}_{e})}{W_{total}}\label{brelec}.
\end{eqnarray}
\section{Numerical results and discussion}
This section will be devoted to the presentation and analysis of the numerical results obtained. We will see exactly how the direction of the laser field can affect the different quantities calculated in the previous section. But, let us first check the consistency of our theoretical calculation by recovering the results previously obtained where the laser field is along the $z$-axis \cite{mouslih}. A laser field propagating in the direction of the $z$-axis is theoretically represented by the wave vector $\mathbf{k}$ with a component along only the $z$-axis: $\mathbf{k}=(0,0,\omega)$. This requires the first polarization vector $\mathbf{a}_1$ to be in the $x$-axis direction and the second $\mathbf{a}_2$ in the $y$-axis direction, so that the three vectors form a direct orthogonal basis ($\mathbf{a}_1,\mathbf{a}_2,\mathbf{k}$). Therefore, we have
\begin{equation}\label{zaxis}
\mbox{\textbf{k} along the \textit{z}-axis}~ 
\begin{cases}
\theta_k=0;~\varphi_k=0~\mbox{or any}&\Longrightarrow ~~~k=(\omega,0,0,\omega),\\
\theta_{a_1}=\pi/2;~\varphi_{a_1}=0&\Longrightarrow ~~~a_1=(0,|\mathbf{a}|,0,0),\\
\theta_{a_2}=\pi/2;~\varphi_{a_2}=\pi/2&\Longrightarrow ~~~a_2=(0,0,|\mathbf{a}|,0).
\end{cases}
\end{equation}
\begin{figure}[hbtp]
\centering
\includegraphics[scale=0.5]{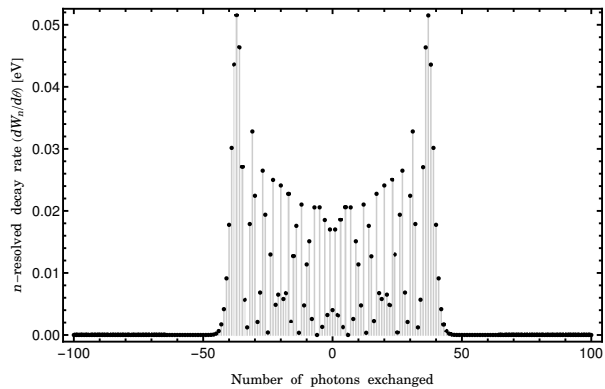}
\caption{The behavior of the $n$-resolved decay rate $dW_{n}/d\theta~(\pi^{-}\rightarrow \mu^{-}\bar{\nu}_{\mu})$ (in units of $10^{-8}$) as a function of the number of photons exchanged $n$ for $\theta=90^{\circ}$. The laser field strength and frequency are $\mathcal{E}_{0}=10^{7}$ V cm$^{-1}$ and $\hbar\omega=1.17~\mbox{eV}$.}\label{env1}
\end{figure}
The result of this particular case is shown in Fig.~\ref{env1}, which illustrates the changes in the $n$-resolved decay rate $(dW_{n}/d\theta)$ in terms of the number of photons exchanged $n$ at field strength $10^{7}$ Vcm$^{-1}$ and frequency $1.17~\mbox{eV}$. It is the same envelope obtained in our previous paper (see Fig.~1(a) in \cite{mouslih}). Thus, the theoretical formalism adopted here is consistent and can lead to all results we have obtained before in \cite{mouslih}.\\
\begin{figure}[h]
 \centering
\includegraphics[scale=0.54]{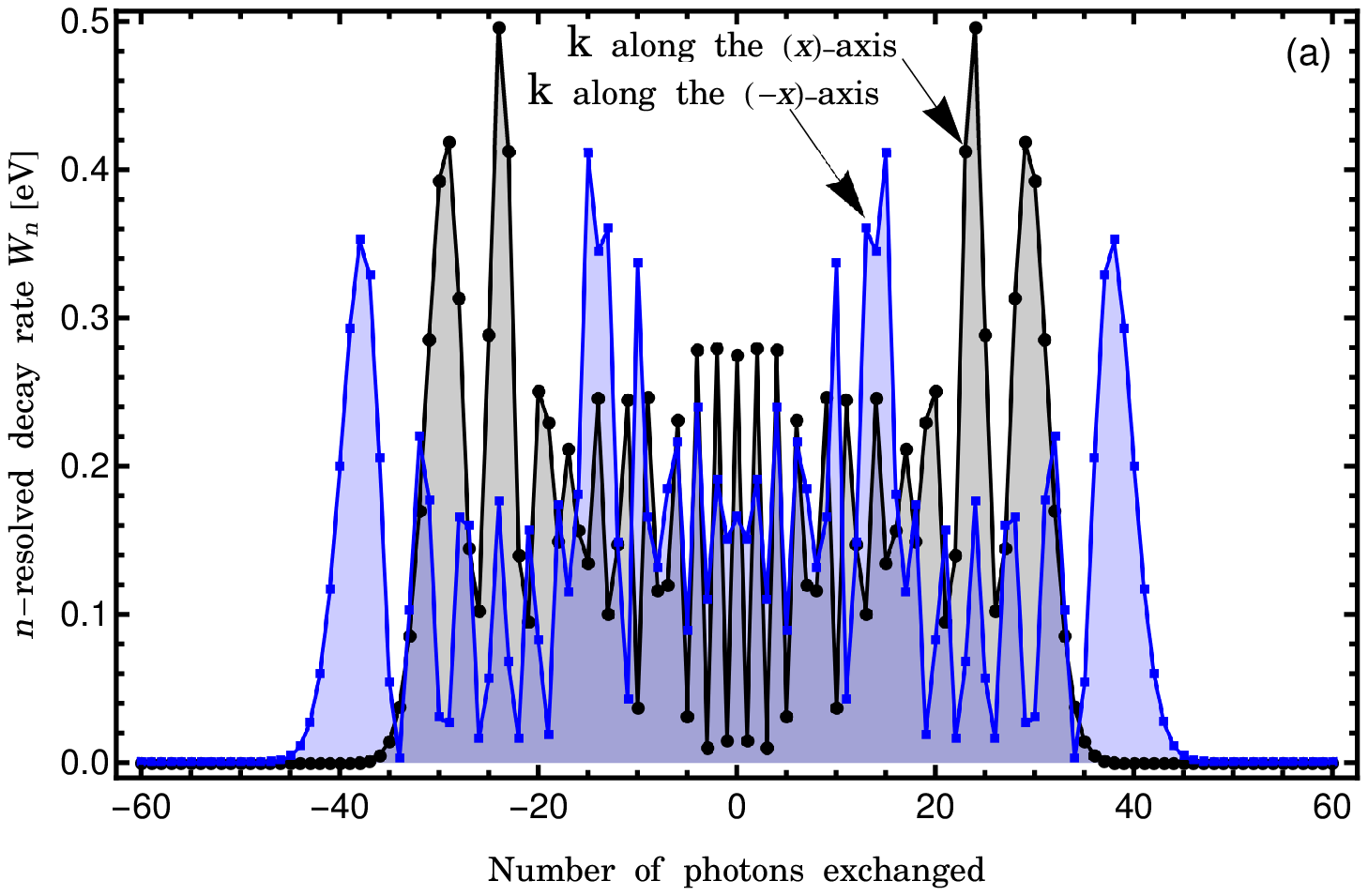}\hspace*{0.2cm}
\includegraphics[scale=0.54]{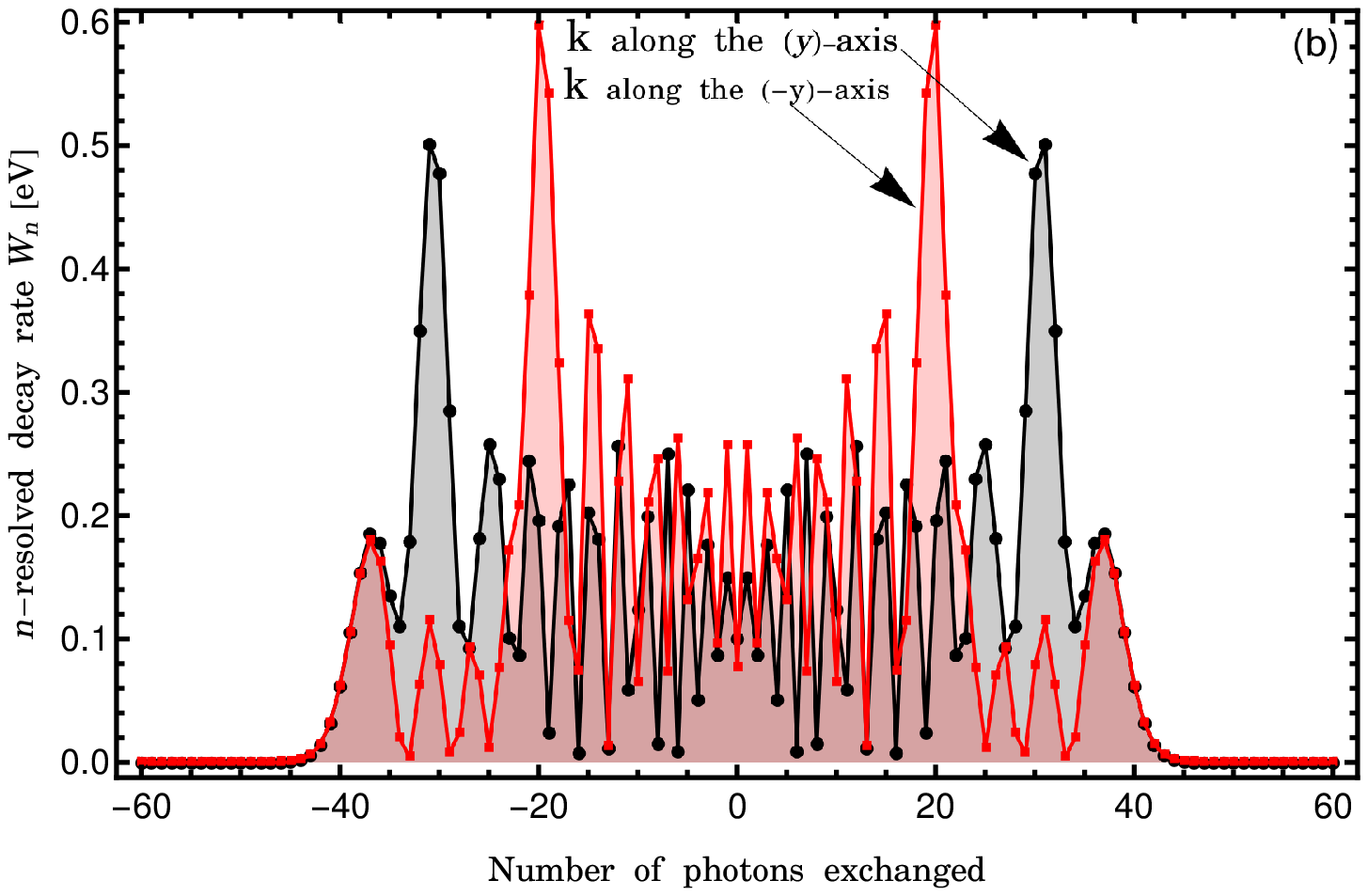}\par\vspace*{0.2cm}
\includegraphics[scale=0.58]{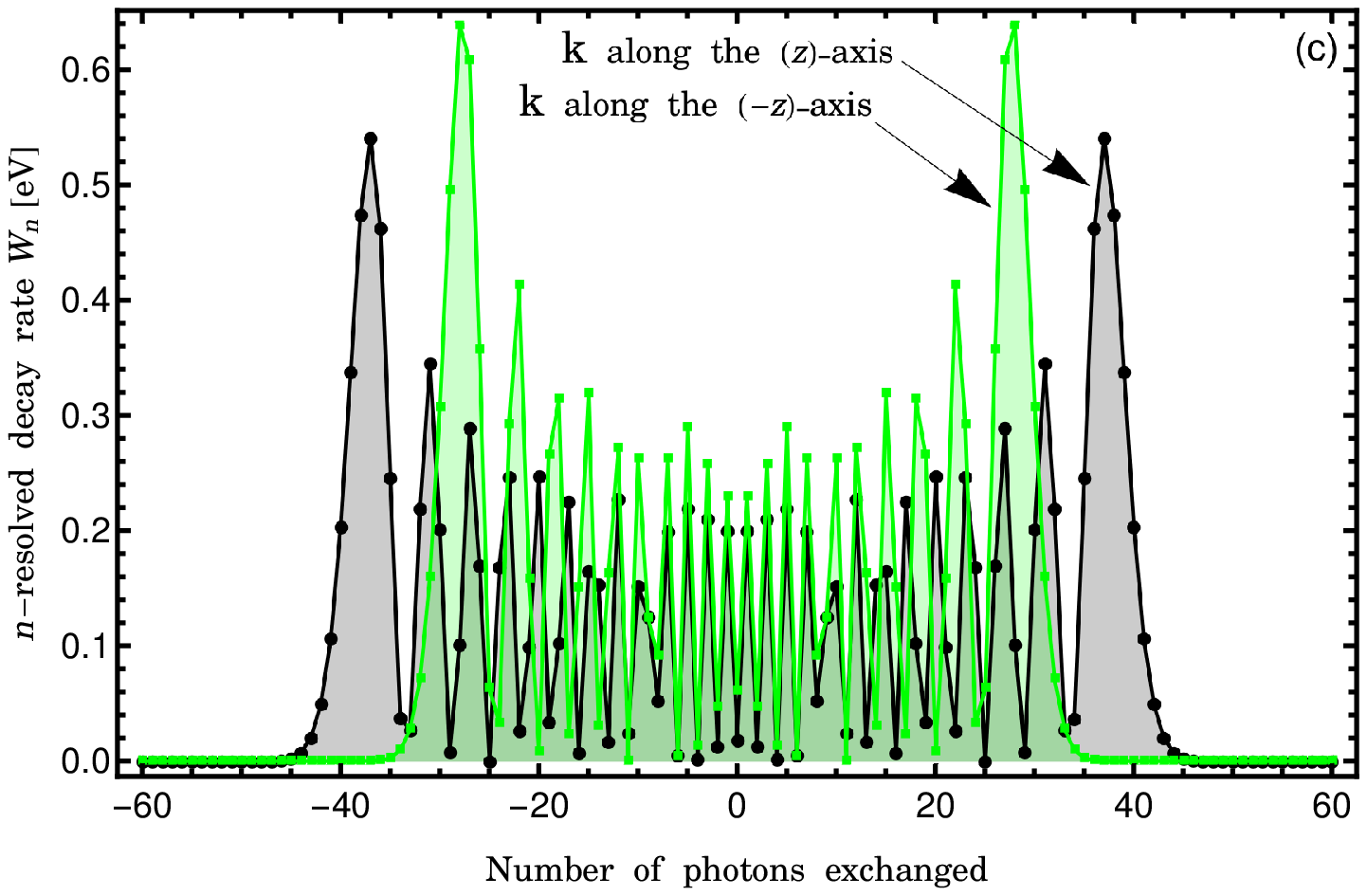}
\caption{The behavior of the total $n$-resolved decay rate $W_{n}(\pi^{-}\rightarrow \mu^{-}\bar{\nu}_{\mu})$ (in units of $10^{-8}$) as a function of the number of photons exchanged $n$ for three laser field directions. The laser field strength and frequency are $\mathcal{E}_{0}=10^{7}$ V cm$^{-1}$ and $\hbar\omega=1.17~\mbox{eV}$.}\label{env2}
\end{figure}
After checking that our calculation is accurate, let us now discuss the effect of the laser field direction on the decay rate. For the $x$ and $y$-axis directions, we set the spherical angles as follows:
\begin{equation}\label{xaxis}
\mbox{\textbf{k} along the \textit{x}-axis}~ 
\begin{cases}
\theta_k=\pi/2;~\varphi_k=0&\Longrightarrow ~~~k=(\omega,\omega,0,0),\\
\theta_{a_1}=\pi/2;~\varphi_{a_1}=\pi/2&\Longrightarrow ~~~a_1=(0,0,|\mathbf{a}|,0),\\
\theta_{a_2}=0;~\varphi_{a_2}=0~\mbox{or any}&\Longrightarrow ~~~a_2=(0,0,0,|\mathbf{a}|).
 \end{cases}
\end{equation}
\begin{equation}\label{yaxis}
 \mbox{\textbf{k} along the \textit{y}-axis}~ 
 \begin{cases}
 \theta_k=\pi/2;~\varphi_k=\pi/2&\Longrightarrow ~~~k=(\omega,0,\omega,0),\\
\theta_{a_1}=0;~\varphi_{a_1}=0~\mbox{or any}&\Longrightarrow ~~~a_1=(0,0,0,|\mathbf{a}|),\\
\theta_{a_2}=\pi/2;~\varphi_{a_2}=0&\Longrightarrow ~~~a_2=(0,|\mathbf{a}|,0,0).\\
\end{cases}
\end{equation}
In Fig.~\ref{env2}, we show the variations of the total $n$-resolved decay rate $W_{n}(\pi^{-}\rightarrow \mu^{-}\bar{\nu}_{\mu})$ (\ref{ws}) (integrated over $d\Omega_{\ell}$) in terms of the number of photons exchanged $n$ for three laser field directions. Each subfigure in Fig.~\ref{env2} contains the changes in $W_{n}$ with respect to a given direction and its opposite. We obtain envelopes that are symmetric with respect to the $n=0$-axis. These envelopes give us information about the photon exchange process (absorption and emission) between the laser field and the decay system. It appears to us through these figures that the order of magnitude varies according to each direction and between each direction and its opposite. Concerning the amount of photons exchanged, we see that it is almost the same ($-40\leq n\leq 40$) for the three directions, but it differs between each axis and its opposite except for the $y$-axis. It can be seen that the number of photons exchanged when the field is along the $y$-axis is the same as along its opposite direction, since the cutoff number is equal to $\pm45$ in both cases (see Fig.~\ref{env2}(b)). As for the $x$ and $z$-axes, there is a slight difference in the number of photons exchanged between the direction and its opposite. For example, along the $x$-axis, the cutoff number is $\pm35$ and along the opposite direction, it is $\pm45$.\\
\begin{table}[h]
\centering
\caption{Values of laser-modified pion lifetime (\ref{pionlifetime}) for different field strengths and with respect to three laser field directions. The laser frequency is $\hbar\omega=1.17~\mbox{eV}$.}
\begin{tabular}{cccc}
%\begin{tabular}{p{5cm}p{5cm}p{6cm}}\multirow{1}{*}{$\mathcal{E}_{0}$ (V cm$^{-1}$)}
 \hline
\multirow{1}{*}{}&\multicolumn{3}{c}{Pion lifetime $\tau_{\pi^{-}}$ (sec)}\\
\cline{2-4} $\mathcal{E}_{0}$ (V cm$^{-1}$)& $\mathbf{k}$ along $x$-axis& $\mathbf{k}$ along $y$-axis & $\mathbf{k}$ along $z$-axis\\
\hline
$10^{1}$ & $2.5419\times 10^{-8}$ & $2.5419\times 10^{-8}$& $2.5419\times10^{-8}$\\
$10^{2}$ & $2.5419\times 10^{-8}$ & $2.5419\times 10^{-8}$ &$2.5419\times10^{-8}$\\
$10^{3}$ & $2.5419\times 10^{-8}$  & $2.541\times 10^{-8}$&$2.5419\times10^{-8}$  \\
$10^{4}$ & $2.5419\times 10^{-8}$ & $2.5419\times 10^{-8}$& $2.5419\times10^{-8}$\\
$10^{5}$ & $2.5419\times 10^{-8}$ & $2.5419\times 10^{-8}$& $2.5419\times10^{-8}$\\
$10^{6}$ & $2.5421\times 10^{-8}$  & $2.542\times 10^{-8}$ & $2.542\times10^{-8}$\\
$10^{7}$ & $5.8895\times 10^{-8}$ &$1.4656\times 10^{-7}$&$9.539\times10^{-8}$ \\
$10^{8}$ & $2.4669\times 10^{-7}$ & $1.4625\times 10^{-6}$&$9.8883\times10^{-7}$ \\
  \hline
\end{tabular}
\label{tab1}
\end{table}
We will now study the variation of other quantities with respect to the different directions of the laser field. We will sum over a specific range of number of photons that we choose to be from $-10$ to $+10$. First, we show in Table~\ref{tab1} the numerical values of the pion lifetime for different field strengths and along three laser field directions. In the absence of the laser, the pion lifetime is equal to $\tau_{\pi^{-}}=(2.6033\pm0.0005)\times 10^{-8}~\mbox{sec}$ \cite{PDG}. From Table~\ref{tab1}, it is clear to us that the laser field at its low strengths (10 to $10^6$ V cm$^{-1}$) remains without significant effect on the lifetime. But, as the laser field strength increases to $10^7$ and $10^8$ V cm$^{-1}$, we notice that the lifetime starts to increase depending on the direction of the laser field. Note that the lifetime increases faster on the $y$ and $z$-axes than on the $x$-axis.\\
\begin{table}[h]
\centering
\caption{Numerical Values of $\mbox{Br}(\pi^{-}\rightarrow \mu^{-}\bar{\nu}_{\mu})$ (\ref{brmuon}) and $\mbox{Br}(\pi^{-}\rightarrow e^{-}\bar{\nu}_{e})$ (\ref{brelec}) for different field strengths and with respect to three laser field directions. The laser frequency is $\hbar\omega=1.17~\mbox{eV}$.}
\begin{tabular}{cccc|ccc}
%\begin{tabular}{p{5cm}p{5cm}p{6cm}}\multirow{1}{*}{$\mathcal{E}_{0}$ (V cm$^{-1}$)}
 \hline
\multirow{1}{*}{}&\multicolumn{3}{c|}{$\mbox{Br}(\pi^{-}\rightarrow \mu^{-}\bar{\nu}_{\mu})$ (\%)}&\multicolumn{3}{c}{$\mbox{Br}(\pi^{-}\rightarrow e^{-}\bar{\nu}_{e})$ (\%)}\\
\cline{2-7} $\mathcal{E}_{0}$ (V cm$^{-1}$)& $\mathbf{k}$ in $x$-axis& $\mathbf{k}$ in $y$-axis & $\mathbf{k}$ in $z$-axis& $\mathbf{k}$ in $x$-axis& $\mathbf{k}$ in $y$-axis & $\mathbf{k}$ in $z$-axis\\
\hline
$10^{1}$ & $99.9874$ & $99.9874$ & $99.9874$&  $0.0126$ & $0.0126$ & $0.0126$    \\
$10^{2}$ & $99.9874$ & $99.9874$ & $99.9874$&  $0.0126$ & $0.0126$ & $0.0126$    \\
$10^{3}$ & $99.9874$ & $99.9874$ & $99.9874$&  $0.0126$ & $0.0126$ & $0.0126$    \\
$10^{4}$ & $99.9874$ & $99.9874$ & $99.9874$&  $0.0126$ & $0.0126$ & $0.0126$    \\
$10^{5}$ & $99.9889$ & $99.9874$ & $99.9875$&  $0.011 $ & $0.0126$ & $0.0124$   \\
$10^{6}$ & $99.9973$ & $99.9939$ & $99.9926$&  $0.0026$ & $0.0096$ & $0.0073$ \\
$10^{7}$ & $99.9994$ & $99.9966$ & $99.9966$&  $0.0005$ & $0.0048$ & $0.0034$\\
$10^{8}$ & $99.9998$ & $99.9968$ & $99.9965$&  $0.0002$ & $0.0097$ & $0.0034$\\
  \hline
\end{tabular}
\label{tab2}
\end{table}
Table~\ref{tab2} contains the values of the two branching ratios for different field strengths. It is well known that the branching ratio of the muon channel is more favored, in the absence of the laser, compared to that of other channels \cite{PDG}. From Table~\ref{tab2}, we note that the laser further enhanced the muon branching ratio and, on the other hand, suppressed the electronic one, which becomes almost nonexistent at high field strengths. The direction of the laser field in this case has no observable effect on the branching ratios. 
\begin{figure}[hbtp]
\centering
\includegraphics[scale=0.6]{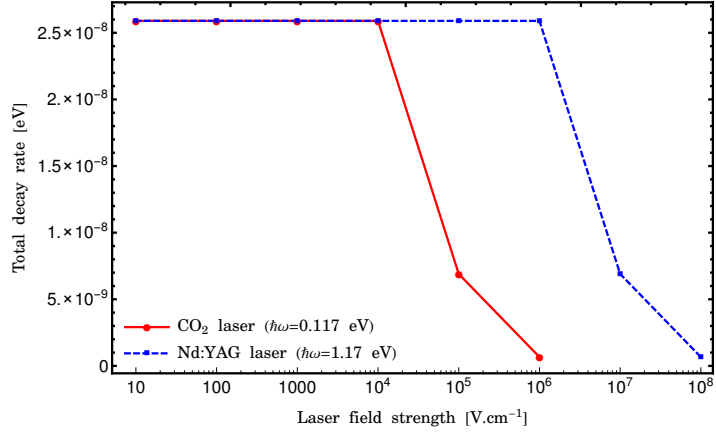}
\caption{Variation of the total decay rate $W_{total}$ as a function of the laser field strength for a Nd:YAG laser ($\hbar\omega=1.17~\text{eV}$) and a $\text{CO}_{2}$ laser ($\hbar\omega=0.117~\text{eV}$). The number of exchanged photons summed over it is $-10\leq n\leq+10$.}\label{fig3}
\end{figure}
After investigating the effect of the laser field direction on different measurable quantities during the pion decay process, let us add some results related to the total decay rate $W_{total}$ defined in Eq.~(\ref{pionlifetime}). To highlight the effect of the laser field strength and frequency on the total decay rate, we show through Fig.~\ref{fig3} the changes of the total decay rate in terms of field strength for two different available frequencies. We note that the total decay rate is not affected by the laser field at low strengths, but it decreases significantly with increasing field strength. Regarding its dependence on the field frequency, it turns out that the laser effect decreases at high frequencies. The effect of $\text{CO}_{2}$ laser begins at strengths above $10^4$ V cm$^{-1}$, but the effect of Nd:YAG laser only appears when we exceed the field strength $10^6$ V cm$^{-1}$. The low-frequency laser affects the total decay rate significantly faster than the high-frequency laser.
\begin{figure}[hbtp]
\centering
\includegraphics[scale=0.64]{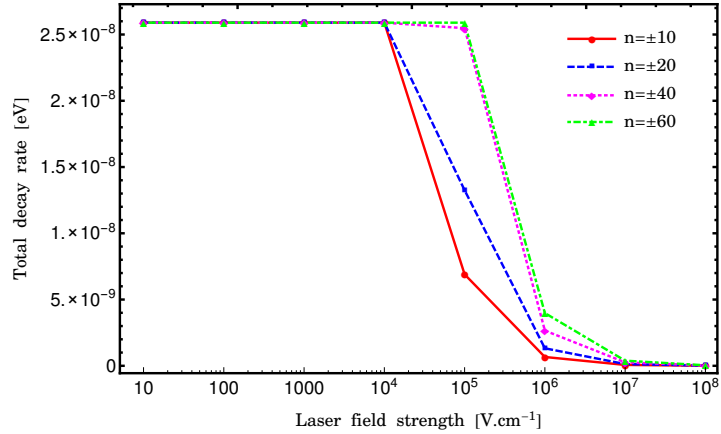}
\caption{The same as in Fig.~\ref{fig3}, but for different numbers of exchanged photons and fixed laser frequency $\hbar\omega=0.117~\text{eV}$. The notation $n=\pm N$ means that $n$ is summed over $-N\leq n\leq +N$.}\label{fig4}
\end{figure}
In Fig.~\ref{fig4}, we plot the variations of the total decay rate as a function of field strength, but now for different numbers of photons exchanged. It is clear that the effect of the laser on the total decay rate is also dependent on the number of photons exchanged, as it gradually decreases with each increase in the number of photons exchanged.
\begin{figure}[hbtp]
\centering
\includegraphics[scale=0.5]{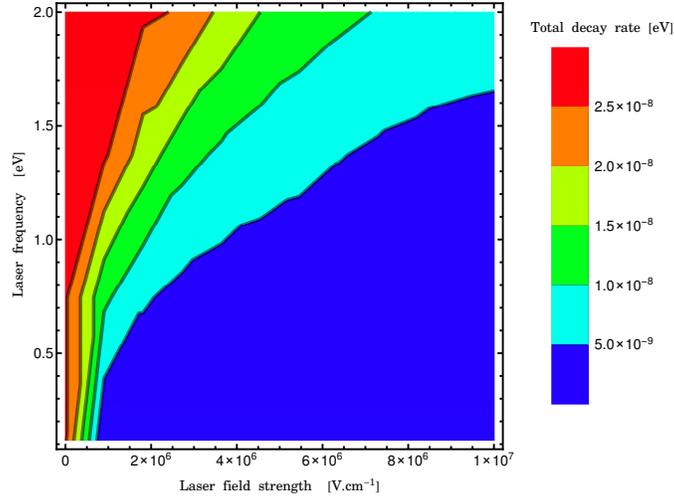}
\caption{The 3-dimensional contour-plot of $W_{total}$ as a function of the laser field strength $\mathcal{E}_{0}$ and frequency $\omega$ for $-5\leq n\leq+5$ exchanged photons.}\label{fig5}
\end{figure}
As an additional graphic illustration, we show in Fig.~\ref{fig5} a 3-dimensional representation of $W_{total}$ as a function of the laser field strength $\mathcal{E}_{0}$ and frequency $\omega$ for $-5\leq n\leq+5$ exchanged photons. This type of plot provides more information and a complete picture of the dependence of $W_{total}$ on the laser parameters. Regarding the variation versus laser frequency, we note that the effect of the laser field decreases at high frequencies, while it becomes significant with increasing field strength at each given frequency. This is perfectly consistent with everything we said above when describing Fig.~\ref{fig3}.
\section{Conclusion}
In this paper, we dealt with the effect of laser field direction on the decay process of pion. We have extended the study of the laser-assisted pion decay process to the case of a general laser field direction. We have concluded that the direction of the laser field does not play an important role in laser-assisted decay processes in which the decaying particle is at rest. The configuration of the laser field direction does not change the nature of the result obtained even if it slightly affects the photon exchange process. This is due to the fact that the decaying particle is at rest in the initial state. Therefore, the geometry of the laser field will inevitably have a significant effect in the case where the initial particle is in motion as in the scattering processes. However, the theoretical formalism presented here seems valid and important as it can be exploited by other researchers for application to scattering processes that occur in the presence of a circularly polarized laser field.
%\section*{References}


\begin{thebibliography}{99}
\bibitem{piazza} A Di Piazza, C M\"{u}ller, K Z Hatsagortsyan and C H Keitel \emph{Rev. Mod. Phys.} \textbf{84} 1177 (2012)
\bibitem{laser} J W Yoon, C Jeon, J Shin, S K Lee, H W Lee, I W Choi, H T Kim, J H Sung and C H Nam \emph{Opt. Express} \textbf{27} 20412 (2019)
\bibitem{manaut1} E Hrour, M El Idrissi, S Taj and B Manaut \emph{Indian J. Phys.} (2021) \url{https://doi.org/10.1007/s12648-021-02087-0} 
\bibitem{manaut2} M El Idrissi, E Hrour, S Taj and B Manaut \emph{Indian J. Phys.} \textbf{95} 2541 (2021)
\bibitem{manaut3} E Hrour, S Taj, A Chahboune, M El Idrissi and B Manaut \emph{Laser Phys.} \textbf{27} 066003 (2017)
\bibitem{manaut4} E Hrour, M El Idrissi, S Taj and B Manaut \emph{Indian J. Phys.} \textbf{89} 783 (2015)
\bibitem{manaut5}  M El Idrissi, S Taj, B Manaut and L Oufni \emph{Indian J. Phys.} \textbf{88} 111 (2014)
\bibitem{Roshchupkin1} A A Lebed' and S P Roshchupkin \emph{Laser Phys. Lett.} \textbf{5} 437 (2008)
\bibitem{Roshchupkin2} S P Roshchupkin and V A Tsybul'nik \emph{Laser Phys. Lett.} \textbf{3} 362 (2006)
\bibitem{Roshchupkin3} S P Roshchupkin, A A Lebed' and E A Padusenko \emph{Laser Phys.} \textbf{22} 1513 (2012)
\bibitem{dahiri} I Dahiri, M Jakha, S Mouslih, B Manaut, S Taj and Y Attaourti \emph{Laser Phys. Lett.} \textbf{18} 096001 (2021)
\bibitem{ouhammou1} M Ouhammou, M Ouali, S Taj and B Manaut \emph{Laser Phys. Lett.} \textbf{18} 076002 (2021)
\bibitem{ouhammou2} M Ouhammou, M Ouali, S Taj and B Manaut \emph{Chin. J. Phys.} (2021) \url{https://doi.org/10.1016/j.cjph.2021.09.012.}
\bibitem{mouha} M Ouali, M Ouhammou, Y Mekaoui, S Taj and B Manaut \emph{Chin. J. Phys.} (2021) \url{https://doi.org/10.1016/j.cjph.2021.10.007}
\bibitem{mouslih} S Mouslih, M Jakha, S Taj, B Manaut and E Siher \emph{Phys. Rev. D} \textbf{102} 073006 (2020)
\bibitem{jakha} M Jakha, S Mouslih, S Taj and B Manaut \emph{Laser Phys. Lett.} \textbf{18} 016002 (2021)
\bibitem{baouahi} M Baouahi, M Ouali, M Jakha, S Mouslih, Y Attaourti, B Manaut, S Taj and R Benbrik \emph{Laser Phys. Lett.} \textbf{18} 106001 (2021)
\bibitem{wdecay} M Jakha, S Mouslih, S Taj, Y Attaourti and B Manaut \emph{Chin. J. Phys.} (2021) \url{https://doi.org/10.1016/j.cjph.2021.09.011}
\bibitem{volkov} D M Volkov \emph{Z. Phys.} \textbf{94} 250 (1935)
\bibitem{landau} V B Berestetskii, E M Lifshitz and L P Pitaevskii \emph{Quantum Electrodynamics} (Oxford U.K.: Butterworth-Heinemann) (1982)
\bibitem{szymanowski} C Szymanowski, V Veniard, R Taieb, A Maquet and C H Keitel \emph{Phys. Rev. A} \textbf{56} 3846 (1997)
\bibitem{greiner} W Greiner and B Muller \textit{Gauge Theory of Weak Interactions} (Berlin: Springer) (2000)
\bibitem{feyncalc} V Shtabovenko, R Mertig and F Orellana \emph{Comput. Phys. Commun.} \textbf{256} 107478 (2020)
\bibitem{PDG} Zyla P A \emph{et al} (Particle Data Group) 2020 \emph{Prog. Theor. Exp. Phys.} \textbf{2020} 083C01
\end{thebibliography}
\end{document}